\documentstyle[aps,preprint,eqsecnum,tighten,prd]{revtex}

\begin{document}
\title{Is there emitted radiation in Unruh effect?}
\author{B. L. Hu}
\address{Department of Physics, University of Maryland\\
College Park, MD 20742, USA\\
E-mail address: hub@physics.umd.edu}
\author{Alpan Raval}
\address{Keck Graduate Institute, 535 Watson Drive\\
Claremont, CA 91711, USA\\
E-mail address: raval@tki.org}
\date{December 1, 2000}
\maketitle

\begin{abstract}
The thermal radiance felt by a uniformly accelerated
detector/oscillator/atom-- the Unruh effect \cite{Unr76} -- is often
mistaken to be some emitted radiation detectable by an
observer/probe/sensor. Here we show by an explicit calculation of the energy
momentum tensor of a quantum scalar field that, at least in 1+1 dimension,
while a polarization cloud is found to exist around the particle trajectory, 
{\it there is no emitted radiation from a uniformly accelerated oscillator
in equilibrium conditions}. Under {\it nonequilibrium} conditions which can
prevail for non-uniformly accelerated trajectories or before the atom or
oscillator reaches equilibrium, there is conceivably radiation emitted, but
that is not what Unruh effect entails.
\end{abstract}

\noindent {\small {\it - Invited talk given by BLH at the Capri Workshop on
Quantum Aspects of Beam Physics, Oct. 2000. To appear in the Proceedings
edited by Pisin Chen (World Scientific, Singapore, 2001)}}

\section{Introduction}

The title question has realistic significance in light of recent
experimental proposals on the detection of `Unruh {\it radiation' emitted}
by linear uniformly accelerated charges\cite{Chen}. Earlier findings of ours 
\cite{RavalPhD,RHA} and others had already addressed this issue, with the
results that, at least in 1+1 dimension model calculations, {\it there is no
emitted radiation from a linear uniformly accelerated oscillator in a steady
state}, even though there exists a polarization cloud around it. There could
be radiation emitted in nonequilibrium conditions, which arise for
non-uniformly accelerated atoms ( for an example of finite time
acceleration, see\cite{RHK}), or during initial transient time for a
uniformly accelerated atom, when its internal states have not yet reached
equilibrium through interaction with the field. For a review of earlier work
on accelerated detectors, see e.g.,\cite{AccDetRev}. For a discussion of
nonequilibrium processes beyond the Unruh effect, see\cite{CapHJ,CapJH}.

After Unruh and Wald's\cite{UnrWal} explication of what a Minkowski observer
sees, Grove\cite{Grove}questioned whether an accelerated detector actually
emits radiated energy. Raine, Sciama and Grove\cite{RSG}(RSG) analyzed what
an inertial observer placed in the forward light cone of the accelerating
detector would measure, and concluded that the detector does not radiate.
Unruh\cite{Unr92}, in an independent calculation, basically concurred with
the findings of RSG but he also showed the existence of extra terms in the
two-point function of the field which could contribute to the excitation of
a detector placed in the forward light cone. Massar, Parantani and Brout\cite
{MPB} (MPB) pointed out that the missing terms in RSG constitute a
polarization cloud around the accelerating detector. Further discussion were
conducted by Hinterleitner\cite{Hin}, Audretsch, M\"{u}ller and Holzmann\cite
{AMH} Massar and Parantani\cite{MPB}.

Both RSG and RHA treated particle- field interaction as a quantum
dissipative system. RSG attributed the lack of radiation from the
accelerated detector to the existence of a fluctuation-dissipation relation
(FDR) governing its dynamics\footnote{%
There is a common misconception that a FDR can be used to explain the
cancellation of radiation reaction by vacuum fluctuations, not realizing
that the former is classical in nature while the latter is a quantum entity.
The FDR in our work exists at the quantum stochastic level and relates the
quantum dissipation in a particle's trajectory \cite{CapJH} or an atom's
internal degrees of freedom \cite{RHA} to the vacuum fluctuations in a
field. It does not involve radiation reaction, which vanishes for a
uniformly accelerated charge because of special conditions existing for the
classical acceleration fields \cite{CapHJ}.}. Using the open system concept
RHA constructed the influence functional and derived a set of coupled
stochastic equations for a system of n-detectors in arbitrary (yet
prescribed) states of motion in a quantum field. One subcase they studied
related to Unruh effect was the influence of an accelerated detector on a
probe (which is not allowed to causally influence the accelerated detector
itself) via the quantum field. They found that most of the terms in the
correlations of the stochastic force acting on the probe cancel each other,
owing to the existence of a correlation-propagation relation, related to the
fluctuation-dissipation relation for the accelerated detector\footnote{%
Such a relation can be equivalently viewed as a construction of the free
field two-point function for each point on either trajectory from the
two-point function along the uniformly accelerated trajectory alone.}. The
remaining terms, which contribute to the excitation of the probe, are shown
to represent correlations of the free field across the future horizon of the
accelerating detector.

Here, we will use the simpler Heisenberg operator method to calculate the
two point function and the energy momentum tensor of a massless quantum
scalar field in a $1+1$ - dimensional Minkowski spacetime minimally coupled
to an accelerated particle with internal oscillator coordinates. Our
analysis (based on Chapter 2 of Alpan Raval's thesis\cite{RavalPhD}) is more
general than that of MPB in that the two-point function is calculated for
the two points lying in arbitrary regions of Minkowski space, and not
restricted to lie to the left of the accelerated oscillator trajectory. We
show where the extra terms in the two point function are which were ignored
in the RSG analysis. More relevant to answering the title question, we show
that at least in two dimensions the energy momentum tensor vanishes
everywhere except on the horizon. This means that beyond the initial
transient, there is no net flux of radiation emitted from the uniformly
accelerated oscillator.

\section{Correlations and Stress Energy of Quantum Field}

\subsection{Minimal coupling particle- field model}

As in RHA, we consider the scalar electrodynamic or ``minimal'' coupling of
oscillators to a scalar field in 1+1 dimensions. This coupling provides a
positive definite Hamiltonian, and is of interest because it resembles the
actual coupling of charged particles to an electromagnetic field. We assume
that the field and the detector are initially decoupled from each other, and
that the field is initially in the Minkowski vacuum state.

The complete action of the minimally coupled field - particle system is 
\begin{eqnarray}
S &=&\frac{1}{2}\int d\tau \left\{ (\frac{dQ}{d\tau })^{2}-\Omega
_{0}^{2}Q^{2}\right\} +e\int d\tau \frac{dQ}{d\tau }\phi (x(\tau ),t(\tau ))
\nonumber \\
&&+\frac{1}{2}\int dx\int dt\left\{ (\frac{\partial \phi }{\partial t})^{2}-(%
\frac{\partial \phi }{\partial x})^{2}\right\} .
\end{eqnarray}
where $\Omega _{0}$ is the bare frequency of the oscillator and $e$ its
coupling constant to the field. Under uniform acceleration, the particle
trajectory parametrized by the proper time $\tau $ is 
\begin{equation}
x(\tau )=a^{-1}\cosh a\tau ;~~~t(\tau )=a^{-1}\sinh a\tau .  \label{2.2}
\end{equation}
Variation of the action leads to the following equations of motion: 
\begin{eqnarray}
\frac{d^{2}Q}{d\tau ^{2}}+\Omega _{0}^{2}Q& =-e\frac{d\phi }{d\tau }(x(\tau
),t(\tau ))  \label{2.3} \\
\frac{\partial ^{2}\phi }{\partial t^{2}}-\frac{\partial ^{2}\phi }{\partial
x^{2}}& =e\int d\tau \frac{dQ}{d\tau }\delta (x-x(\tau ))\delta (t-t(\tau )).
\label{2.4}
\end{eqnarray}
Because the action is a quadratic functional of the field and oscillator
variables, these are also the Heisenberg operator equations of motion for
the system. We shall thus view the above equations as operator equations
from now on.

The field equations are solved by introducing the retarded Green function of
a massless scalar field in $1+1$ dimensions: 
\begin{equation}
\phi(x,t) = \phi_{0} (x,t) + e \int_{-\infty}^{\infty} d\tau\frac{dQ}{d\tau}
G_{ret} (x,t ; x(\tau),t(\tau))  \label{2.5}
\end{equation}
where $\phi_{0}$ is a solution to the homogenous field equations
corresponding to $Q=0$. We will find it convenient to introduce the null
coordinates $u=t-x$ and $v=t+x$. Correspondingly, we also find it convenient
to define the regions $F,P,R$ and $L$ of Minkowski space as ($R$ is called
the Rindler wedge)

\begin{eqnarray}
F:u>0, v>0 & ~~~ P:u<0, v<0  \nonumber \\
R:u<0, v>0 & ~~~ L:u>0, v<0 .  \label{2.6}
\end{eqnarray}
In terms of the $(u,v)$ coordinates, the retarded Green function for a
massless scalar field in $1+1$ dimensions takes the form: 
\begin{eqnarray}
G_{ret} (x,t ; x(\tau),t(\tau)) & = \frac{1}{2} \theta(t-t(\tau)-x+x(\tau))
\theta(t-t(\tau) +x-x(\tau))  \nonumber \\
& = \frac{1}{2} \theta(u+a^{-1}e^{-a\tau}) \theta(v-a^{-1} e^{a\tau }).
\label{2.7}
\end{eqnarray}
With this substitution, an integration by parts in Eq. (\ref{2.5}) yields: 
\begin{eqnarray}
\phi(x,t) & = \phi_{0} (x,t) + \frac{e}{2} \left[ \theta(-u) \theta
(-\lambda) Q(-a^{-1}ln(\mid au\mid))\right.  \nonumber \\
& + \left. \theta(v) \theta(\lambda) Q(a^{-1}ln(\mid av\mid)) \right]
\label{2.8}
\end{eqnarray}
where we have also defined $\lambda=1+a^{2}uv$. The oscillator trajectory
satisfies $\lambda=0$. The quantities $-a^{-1}ln(\mid au\mid)$ and $%
a^{-1}ln(\mid av\mid)$ are just the retarded times of the point $(x,t)$,
according to whether it lies to the right or the left of the accelerated
trajectory, respectively. These two cases are distinguished by the
appearance of the step functions with argument $\mp\lambda$. The step
functions in $u$ and $v$ distinguish the cases when the point lies anywhere
in the past light cone or anywhere in the forward light cone of the
accelerated particle (these two conditions are simultaneously satisfied only
in the Rindler wedge). With this in mind, we see that the first term linear
in the coupling constant contributes only for points to the right of the
oscillator trajectory, whereas the second term contributes only for points
to the left of the oscillator trajectory and within the forward light cone
of the oscillator. In particular, as expected, there is no correction to the
field operator in the region $L\cup P$, which cannot be causally influenced
by the accelerated trajectory.

Along the accelerated trajectory, the solution for $\phi$ reduces to 
\begin{equation}
\phi(x(\tau),t(\tau)) = \phi_{0} (x(\tau),t(\tau)) + \frac{e}{2} Q(\tau).
\end{equation}
Putting this back to the equation of motion for $Q$, (\ref{2.3}), we obtain: 
\begin{equation}
\frac{d^{2} Q}{d\tau^{2}} + \frac{e^{2}}{2} \frac{dQ}{d\tau} +
\Omega_{0}^{2} Q = -e \frac{d\phi_{0} }{d\tau}(x(\tau),t(\tau)).
\label{2.10}
\end{equation}
The term linear in the proper velocity of the oscillator degree of freedom
arises from the oscillator - field interaction and corresponds to
dissipation of a quantum origin in the oscillator.

\subsection{Equation of motion for the detector}

The above equation of motion is easily solved. If the oscillator field
interaction has always been switched on, the oscillator has reached a steady
state at any finite time. We can then ignore transient terms in the solution
for $Q$ and obtain: 
\begin{equation}
Q(\tau) = -\frac{e}{\Omega} \int_{-\infty}^{\tau} d\tau^{\prime}\sin
\Omega(\tau-\tau^{\prime}) e^{-\gamma(\tau-\tau^{\prime})} \frac{d\phi_{0} }{%
d\tau^{\prime}}(x(\tau^{\prime}),t(\tau^{\prime}))
\end{equation}
where we have defined the dissipation constant $\gamma=\frac{e^{2}}{4}$, and
the frequency $\Omega=\sqrt{\Omega_{0}^{2} -\gamma^{2}}$. We may also solve
equation (\ref{2.10}) in frequency space. Ignoring transients as before, we
obtain 
\begin{equation}
\tilde{Q}(\omega)\equiv\frac{1}{2\pi}\int_{-\infty}^{\infty}d\tau
e^{-i\omega\tau}Q(\tau)=\chi_{\omega}\tilde{J}(\omega)
\end{equation}
where 
\begin{equation}
\tilde{J}(\omega)=-\frac{e}{2\pi}\int_{-\infty}^{\infty}d\tau
e^{-i\omega\tau }\phi_{0}(x(\tau),t(\tau))
\end{equation}
and $\chi_{\omega}$ is the impedance function of the oscillator, given by 
\begin{equation}
\chi_{\omega}=i\omega(-\omega^{2}+\Omega_{0}^{2}+2i\omega\gamma)^{-1}.
\end{equation}
It satisfies the identity 
\begin{equation}
\chi_{\omega}+\chi_{\omega}^{\ast}=4\gamma\mid\chi_{\omega}\mid^{2}
\label{2.15}
\end{equation}
which has the form of a fluctuation- dissipation relation. We now expand $%
\phi_{0}$ in Minkowski normal modes: 
\begin{equation}
\phi_{0}(t,x)=\phi_{0}^{(+)}+\phi_{0}^{(-)}=\int_{-\infty}^{\infty}\frac {%
d^{2}k}{\sqrt{(2\pi)^{2}2\omega_{k}}}{\bf a_{k}}e^{i(kx-\omega_{k}t)}+h.c.
\label{2.16}
\end{equation}
where $h.c.$ denotes Hermitian conjugate and $\phi_{0}^{(-)}$ is the
Hermitian conjugate of $\phi_{0}^{(+)}$. The operators ${\bf {a_{k}}}$
annihilate the Minkowski vacuum. Based on this separation of the field into
positive and negative frequency parts, we obtain the corresponding
separation of the oscillator degree of freedom: 
\begin{equation}
Q(\tau)=Q^{(+)}(\tau)+Q^{(-)}(\tau)
\end{equation}
where 
\begin{equation}
Q^{(+)}(\tau)=-\frac{e}{\Omega}\int_{-\infty}^{\tau}d\tau^{\prime}\sin
\Omega(\tau-\tau^{\prime})e^{-\gamma(\tau-\tau^{\prime})}\frac{d\phi_{0}}{%
d\tau^{\prime}}^{(+)}(x(\tau^{\prime}),t(\tau^{\prime}))  \label{2.18}
\end{equation}
and $Q^{(-)}$ is the Hermitian conjugate of $Q^{(+)}$.

On the accelerated trajectory, Eq. (\ref{2.16}) gives 
\begin{equation}
\phi _{0}^{(+)}(x(\tau ),t(\tau ))=\int_{-\infty }^{\infty }\frac{d^{2}k}{%
\sqrt{(2\pi )^{2}2\omega _{k}}}{\bf a_{k}}[e^{\frac{ik}{a}e^{-a\tau }}\theta
(k)+e^{\frac{ik}{a}e^{a\tau }}\theta (-k)].
\end{equation}
Introducing the Fourier transforms of $e^{\frac{ik}{a}e^{-a\tau }}$ and $e^{%
\frac{ik}{a}e^{a\tau }}$, we get 
\begin{eqnarray}
\phi _{0}^{(+)}(x(\tau ),t(\tau ))& =\frac{1}{2\pi a}\int_{-\infty }^{\infty
}\frac{d^{2}k}{\sqrt{(2\pi )^{2}2\omega _{k}}}{\bf a_{k}}\int_{-\infty
}^{\infty }d\omega e^{-i\omega \tau }e^{\frac{\pi \omega }{2a}}\times 
\nonumber \\
& \left[ \Gamma (-\frac{i\omega }{a})\mid \frac{k}{a}\mid ^{\frac{i\omega }{a%
}}\theta (k)+\Gamma (\frac{i\omega }{a})\mid \frac{k}{a}\mid ^{-\frac{%
i\omega }{a}}\theta (-k)\right] .
\end{eqnarray}
Differentiating with respect to $\tau $ and substituting in the equation for 
$Q^{(+)}(\tau )$, (\ref{2.18}), we obtain, after carrying out the
integration over $\tau $, an expression of $Q^{(+)}(\tau )$ . Then we can
substitute it back into the equation for the field operator (\ref{2.8}) and
get 
\begin{eqnarray}
\phi _{int}^{(+)}(x,t) &=&-\frac{\gamma }{\pi a}\int_{-\infty }^{\infty }%
\frac{d^{2}k}{\sqrt{(2\pi )^{2}2\omega _{k}}}{\bf a_{k}}\int_{-\infty
}^{\infty }d\omega e^{\frac{\pi \omega }{2a}}\chi _{\omega }^{\ast } 
\nonumber \\
&&\times \left[ \Gamma (-\frac{i\omega }{a})\mid \frac{k}{a}\mid ^{\frac{%
i\omega }{a}}\theta (k)+\Gamma (\frac{i\omega }{a})\mid \frac{k}{a}\mid ^{-%
\frac{i\omega }{a}}\theta (-k)\right]  \nonumber \\
&&\times \left[ \mid au\mid ^{\frac{i\omega }{a}}\theta (-u)\theta (-\lambda
)+\mid av\mid ^{-\frac{i\omega }{a}}\theta (v)\theta (\lambda )\right]
\end{eqnarray}
where $\phi _{int}^{(+)}(x,t)=\phi (x,t)-\phi _{0}^{(+)}(x,t)$ accounts for
the interaction of the quantum field with the oscillator.

\subsection{Two point function of the field}

In order to evaluate the two-point function $\langle \phi (x,t)\phi
(x^{\prime },t^{\prime })\rangle $ in the Minkowski vacuum, we first
recognize that it is equal to $\langle \phi ^{(+)}(x,t)\phi ^{(-)}(x^{\prime
},t^{\prime })\rangle $. This is because only operator products of the form $%
{\bf a_{k}}{\bf a_{k}}^{\dagger }$ contribute when taking the expectation
value in the Minkowski vacuum. Denoting $\langle \phi (x,t)\phi (x^{\prime
},t^{\prime })\rangle $ by $G(x,t;x^{\prime },t^{\prime })$ and $\langle
\phi _{0}(x,t)\phi _{0}(x^{\prime },t^{\prime })\rangle $ by $%
G_{f}(x,t;x^{\prime },t^{\prime })$, we obtain 
\begin{eqnarray}
G(x,t;x^{\prime },t^{\prime })-G_{f}(x,t;x^{\prime },t^{\prime }) &=&\langle
\phi _{0}^{(+)}(x,t)\phi _{int}^{(-)}(x^{\prime },t^{\prime })\rangle  
\nonumber \\
+\langle \phi _{int}^{(+)}(x,t)\phi _{0}^{(-)}(x^{\prime },t^{\prime
})\rangle  &&+\langle \phi _{int}^{(+)}(x,t)\phi _{int}^{(-)}(x^{\prime
},t^{\prime })\rangle .
\end{eqnarray}
Using the expression for $\phi _{0}^{(+)}(x,t)$ and $\phi
_{int}^{(-)}(x^{\prime },t^{\prime })$ we obtain 
\begin{eqnarray}
\lefteqn{G(x,t;x^{\prime },t^{\prime })-G_{f}(x,t;x^{\prime },t^{\prime }=-%
\frac{\gamma }{2\pi }\int_{-\infty }^{\infty }\frac{d\omega }{\omega }(1-e^{-%
\frac{2\pi \omega }{a}})^{-1}\times }  \nonumber \\
&&\left[ \mid \frac{au}{au^{\prime }}\mid ^{\frac{i\omega }{a}}\theta
(-u)\theta (-u^{\prime })\left\{ \chi _{\omega }^{\ast }\theta (-\lambda
)+\chi _{\omega }\theta (-\lambda ^{\prime })-4\gamma \mid \chi _{\omega
}\mid ^{2}\theta (-\lambda )\theta (-\lambda ^{\prime })\right\} \right.  
\nonumber \\
+ &\mid &\frac{av}{av^{\prime }}\mid ^{-\frac{i\omega }{a}}\theta (v)\theta
(v^{\prime })\left\{ \chi _{\omega }^{\ast }\theta (\lambda )+\chi _{\omega
}\theta (\lambda ^{\prime })-4\gamma \mid \chi _{\omega }\mid ^{2}\theta
(\lambda )\theta (\lambda ^{\prime })\right\}   \nonumber \\
+ &\mid &a^{2}uv^{\prime }\mid ^{\frac{i\omega }{a}}\theta (-u)\theta
(v^{\prime })\left\{ \chi _{\omega }^{\ast }\theta (-\lambda )+\chi _{\omega
}\theta (\lambda ^{\prime })-4\gamma \mid \chi _{\omega }\mid ^{2}\theta
(-\lambda )\theta (\lambda ^{\prime })\right\}   \nonumber \\
&&+\left. \mid a^{2}u^{\prime }v\mid ^{-\frac{i\omega }{a}}\theta
(-u^{\prime })\theta (v)\left\{ \chi _{\omega }^{\ast }\theta (\lambda
)+\chi _{\omega }\theta (-\lambda ^{\prime })-4\gamma \mid \chi _{\omega
}\mid ^{2}\theta (\lambda )\theta (-\lambda ^{\prime })\right\} \right]  
\nonumber \\
&&-\frac{\gamma }{4\pi }\int_{-\infty }^{\infty }\frac{d\omega }{\omega }%
(\sinh \frac{\pi \omega }{a})^{-1}\times   \nonumber \\
\lbrack  &\mid &\frac{au}{au^{\prime }}\mid ^{\frac{i\omega }{a}}\left\{
\chi _{\omega }^{\ast }\theta (-u)\theta (-\lambda )\theta (u^{\prime
})+\chi _{\omega }\theta (-u^{\prime })\theta (-\lambda ^{\prime })\theta
(u)\right\}   \nonumber \\
+ &\mid &\frac{av}{av^{\prime }}\mid ^{-\frac{i\omega }{a}}\left\{ \chi
_{\omega }^{\ast }\theta (v)\theta (\lambda )\theta (-v^{\prime })+\chi
_{\omega }\theta (v^{\prime })\theta (\lambda ^{\prime })\theta (-v)\right\} 
\nonumber \\
+ &\mid &a^{2}uv^{\prime }\mid ^{\frac{i\omega }{a}}\left\{ \chi _{\omega
}^{\ast }\theta (-u)\theta (-\lambda )\theta (-v^{\prime })+\chi _{\omega
}\theta (v^{\prime })\theta (\lambda ^{\prime })\theta (u)\right\}  
\nonumber \\
+ &\mid &a^{2}u^{\prime }v\mid ^{-\frac{i\omega }{a}}\left\{ \chi _{\omega
}^{\ast }\theta (u^{\prime })\theta (\lambda )\theta (v)+\chi _{\omega
}\theta (-u^{\prime })\theta (-\lambda ^{\prime })\theta (-v)\right\} ].
\label{2.25}
\end{eqnarray}
The role of the relation (\ref{2.15}) in the cancellation of various terms
in the first half of the above expression is thus made explicit. Different
terms will vanish depending on which region of Minkowski space each of the
two points is in, and according to whether these points are to the left or
the right of the accelerated trajectory.

The above result can also be obtained via a different quantization
procedure. Instead of expanding the field in Minkowski modes, as above, we
can use Unruh modes which are linear combinations of Rindler modes and
positive frequency with respect to Minkowski time (see, for example\cite
{BirDav}). These modes are easier to handle in the manipulations involved.
However, they have the disadvantage of being defined differently in each
region ($F,P,R$ and $L$). Although the Rindler modes are defined only in $R$
and $L$, the Unruh modes, as linear combinations of Rindler modes, can be
analytically extended to the entire spacetime. One then computes the two
point function in a desired region by expanding the field in a complete set
of Unruh modes as defined by analytic extension to that region. Of course,
one always needs the mode decomposition in $R$, because the field operator
at an arbitrary point depends both on the free field operator at that point
as well as on the accelerated trajectory, which lies in $R$ (see equations (%
\ref{2.8}) and (\ref{2.18})). This procedure will not be repeated here, as
it leads to the same result.

\subsection{Energy Momentum Tensor}

Let us first consider the coincidence limit of the two point function. In
that case all terms involving $u$ and $u^{\prime }$ or $v$ and $v^{\prime }$
vanish as a consequence of the relation (\ref{2.15}). The remaining terms
can be simplified to give: 
\begin{eqnarray}
\lefteqn{\langle \varphi ^{2}(x,t)\rangle -\langle \varphi
_{0}^{2}(x,t)\rangle =-\frac{\gamma }{2\pi }q(v)\int_{-\infty }^{\infty }%
\frac{d\omega }{\omega }(1-e^{-\frac{2\pi \omega }{a}})^{-1}\times } 
\nonumber \\
& [\mid a^{2}uv\mid ^{\frac{i\omega }{a}}\{\chi _{\omega }^{\ast }\theta
(-u)\theta (-\lambda )+\chi _{\omega }\theta (\lambda )(\theta (u)e^{-\frac{%
\pi \omega }{a}}+\theta (-u))\}  \nonumber \\
& +\mid a^{2}uv\mid ^{-\frac{i\omega }{a}}\{\chi _{\omega }\theta (-u)\theta
(-\lambda )+\chi _{\omega }^{\ast }\theta (\lambda )(\theta (u)e^{-\frac{\pi
\omega }{a}}+\theta (-u))\}].
\end{eqnarray}
This corresponds to a static polarization cloud confined to the region $%
F\cup R$, i.e. $v>0$. It is static because it is a function of $%
uv=t^{2}-x^{2}$ in each region. Thus it is constant along any accelerated
world line in particular. In $F$, the curves $t^{2}-x^{2}=constant$ are
spacelike curves and therefore do not correspond to world-lines of physical
particles. Therefore any physical detector in $F$ will respond to the field
in a non-trivial, time-dependent way.

However, it is simple to show that the renormalized energy-momentum tensor
of the field vanishes everywhere except at the past null horizon $v=0$ of
the accelerated trajectory, and on the accelerated trajectory itself. The
energy-momentum tensor is renormalized by subtracting out the free field
contribution. It is thus given by 
\begin{eqnarray}
T_{uu} &=&\lim_{u^{\prime }\rightarrow u,v^{\prime }\rightarrow v}\partial
_{u}\partial _{u^{\prime }}(G(x,t;x^{\prime },t^{\prime
})-G_{f}(x,t;x^{\prime },t^{\prime }))  \nonumber \\
T_{vv} &=&\lim_{u^{\prime }\rightarrow u,v^{\prime }\rightarrow v}\partial
_{v}\partial _{v^{\prime }}(G(x,t;x^{\prime },t^{\prime
})-G_{f}(x,t;x^{\prime },t^{\prime }))  \nonumber \\
T_{uv} &=&0.
\end{eqnarray}

Going back to the expression (\ref{2.25}) for the two-point function, we
find, in $P\cup L$ (i.e. $v,v^{\prime }<0$), that $G(x,t;x^{\prime
},t^{\prime })-G_{f}(x,t;x^{\prime },t^{\prime })=0$. Thus the renormalized
energy momentum tensor trivially vanishes there. In the region $F\cup R$,
and to the left of the trajectory, $\lambda ,\lambda ^{\prime }>0$, we have 
\begin{eqnarray}
G\left( x,t;x^{\prime },t^{\prime }\right) -G_{f}\left( x,t;x^{\prime
}t^{\prime }\right)  &=&-\frac{\gamma }{2\pi }\int_{-\infty }^{\infty }\frac{%
d\omega }{\omega }\left( 1-e^{-\frac{2\pi \omega }{a}}\right) ^{-1} \\
&&\times \lbrack \left| a^{2}uv^{\prime }\right| ^{\frac{i\omega }{a}}\chi
_{\omega }\left( \theta \left( -u\right) +\theta \left( u\right) e^{-\frac{%
\pi \omega }{a}}\right)   \nonumber \\
&&+\left| a^{2}u^{\prime }v\right| ^{-\frac{i\omega }{a}}\chi _{\omega
}^{\ast }\left( \theta \left( -u^{\prime }\right) +\theta \left( u^{\prime
}\right) e^{-\frac{\pi \omega }{a}}\right) ].  \nonumber
\end{eqnarray}
The terms involving $u$, $u^{\prime }$ and $v$, $v^{\prime }$ all vanish as
a consequence of (\ref{2.15}). The remaining cross-terms do not contribute
to the energy-momentum tensor, as can be checked by straightforward
differentiation.

Similarly, to the right of the trajectory, ($\lambda ,\lambda ^{\prime }<0$%
), we obtain 
\begin{eqnarray}
\langle \phi (x,t)\phi (x^{\prime },t^{\prime })\rangle  &-\langle \phi
_{0}(x,t)\phi _{0}(x^{\prime },t^{\prime })\rangle =&-\frac{\gamma }{2\pi }%
\int_{-\infty }^{\infty }\frac{d\omega }{\omega }(1-e^{-\frac{2\pi \omega }{a%
}})^{-1} \\
&&\times \left| a^{2}uv^{\prime }\right| ^{\frac{i\omega }{a}}\chi _{\omega
}^{\ast }+\left| a^{2}u^{\prime }v\right| ^{-\frac{i\omega }{a}}\chi
_{\omega }].
\end{eqnarray}
The energy-momentum tensor vanishes here as well, in a similar fashion.

If we therefore consider a world-tube formed by two accelerated world-lines
with $\lambda>0$ and $\lambda<0$ in the Rindler wedge, then this tube
encloses the accelerated trajectory $\lambda=0$. Also the energy-momentum
tensor vanishes everywhere on the boundary of the tube. Hence there is no
flux of energy-momentum, or radiation from the oscillator at $\lambda=0$.

The cross-terms in $u,v^{\prime}$ and $u^{\prime},v$ which appear in the
above expressions are missing in RSG. Although we have found that they do
not contribute to the energy-momentum tensor, they do signal the presence of
a polarization cloud around the oscillator. These results support those of
Unruh\cite{Unr92}and MPB\cite{MPB}. However, the above analysis has the
advantage of clearly displaying the role of the ``fluctuation-dissipation
relation'' (\ref{2.15}) in the cancellation of terms which would naively be
expected to contribute to the energy-momentum. Also, we have here computed
an expression for the two-point function which is valid over the entire
spacetime. This is a generalization of previous work.

Calculation is underway in four dimensional spacetime, which is certainly
more physical. In this case\cite{TRH} we expect to see the ordinary
classical radiation of the Larmor type from a uniformly accelerated charge,
but the question of interest to us is whether in 4D there is emitted
radiation. If there were it should manifest in the content of the energy
momentum tensor and, being of quantum nature, discernible from the classical
radiation. This would further clarify any existing confusion on the nature
of Unruh radiation. \newline

\section*{Acknowledgment}

We thank Pisin Chen for his invitation to this interesting workshop and
Stefania Petracca for her warm hospitality. This research is supported in
part by NSF grant PHY98-00967

\end{document}